\documentstyle[
	aps,
	twocolumn, 
        twocolumn
	]{revtex}

\begin{document}


\title{
Pattern Formation and a Clustering Transition 
in Power-Law Sequential Adsorption
}
\author{
Ofer Biham$^{1}$, Ofer Malcai$^{1}$, 
Daniel A. Lidar (Hamburger)$^{2}$ and David Avnir$^{3}$
}
\address{
$^{1}$
Racah Institute of Physics, The Hebrew University, Jerusalem
91904, Israel
}
\address{
$^{2}$
Department of Chemistry, University of California,
Berkeley, CA 94720, USA
}
\address{
$^{3}$
Institute of Chemistry, The Hebrew University, 
Jerusalem 91904, Israel
}
\maketitle

\begin{abstract}
\newline{}
A new model that describes adsorption and clustering
of particles on
a surface is introduced.
A {\it clustering} transition is found 
which separates between a phase of weakly correlated 
particle distributions and a phase of strongly correlated
distributions in which the particles 
form localized fractal clusters.
The order parameter of the transition is identified
and the fractal nature of both phases is examined.
The model is relevant to a large class of clustering
phenomena such as aggregation and growth on surfaces,
population distribution in cities, plant and bacterial 
colonies as well as gravitational clustering.
\end{abstract}

\pacs{64.60.Ak,61.43.Hv,82.20.Mj,68.55.-a}

Many of the growth and pattern formation phenomena in nature
occur via adsorption and clustering
of particles on surfaces
\cite{Barabasi95,Zhang97}.
The richness of these phenomena may be attributed to the
great variety of 
structures and symmetries of the adsorbed particles and substrates.
Nonequilibrium growth models often
give rise to fractal structures, which are statistically
self similar over a range of length-scales
\cite{Mandelbrot82}.
In a large class of surface adsorption systems, 
the dominant dynamical process is the 
{\it diffusion} of 
the adsorbed 
particles 
which hop randomly on the surface 
until they nucleate into
immobile clusters 
\cite{Zhang97}.
The formation of {\it fractal clusters},
which are typical in these systems can be
described by the diffusion limited aggregation (DLA) process
\cite{Witten81}.
In DLA a cluster of particles grows due to a slow flux
of particles which diffuse as random walkers until they attach to
the cluster.
The model describes a great variety of aggregation processes
such as 
island growth
in molecular beam epitaxy
\cite{Zhang97},
electrodeposition,
viscous fingering,
dielectric breakdown,
and various biological systems
\cite{Meakin88}.
In many other physical systems, once
an adsorbed particle sticks to the surface it becomes
immobile.
These systems can be described 
by random sequential adsorption (RSA) processes
\cite{Evans93}.
Within the RSA processes, 
one should distinguish between systems in which
particles cannot overlap and systems in which they can adsorb
on top of each other. 
Systems in which particles cannot overlap 
tend to reach a jamming limit, in which the sticking
probability of a new particle vanishes
\cite{Viot92}.
Models which allow multilayer growth 
describe a large class of physical systems, including
deposition of colloids, liquid crystals
\cite{Chick93}, 
polymers and fiber particles
\cite{Provatas95,Provatas97}.
Recently, the case of power law distribution of particle sizes
was studied both for uncorrelated adsorption
\cite{Hermann91}
and for non-overlapping particles
\cite{Brilliantov96}.
In the case of uncorrelated adsorption it was found that
the boundary of the particle clusters 
is fractal
\cite{Hermann91}.
For non-overlapping particles, it was found that the
area that remains exposed is fractal
\cite{Brilliantov96}.

Models that describe growth dynamics 
have been employed 
in recent years in a vast range of scientific fields
as diverse as city organization and growth 
\cite{Batty94,Makse95},
city and highway traffic 
\cite{Wolf95}
and growth of bacterial colonies
\cite{Ben-Jacob92}.
A common feature is the
tendency of the basic objects to form clusters of
high density (typically of fractal shape), 
surrounded by low density areas or voids.
Other examples of clustering appear in the distribution of
mass in the universe
\cite{Peebles93},
in dissipative gases
and granular flow
\cite{Goldhirsch93}
as well as in step bunching on
crystal surfaces during growth
\cite{Kandel94}
and due to electromigration
\cite{Yang96}.
The phenomenon of cluster formation is therefore generic
in a broad class of systems in spite of the fact that the
pattern forming dynamical processes may vary substantially from system
to system. This richness of clustering phenomena is not yet fully backed-up
by appropriate models.

In this Letter we introduce a new model: the
Power-Law Sequential Adsorption (PLSA) model
which describes a variety of surface adsorption 
and clustering processes.
This model 
leads to a rich variety of structures, many of which are
fractal, which mimic the experimental morphologies
found in the examples cited above.
In particular, it exhibits a unique
{\it clustering} transition which 
separates between a weakly correlated phase in which the
adsorbed particles are distributed homogeneously on the
surface and a strongly correlated phase in which they 
form clusters.  

In the PLSA model
circular particles of diameter $d$
are randomly deposited on a two dimensional (2D) 
surface one at a time. 
The deposition process starts from an initial state where
there is one seed particle on the surface.
The sticking probability 
$0 \le p \le 1$
of a newly deposited particle is determined
by the distance $r$ from its center to the
center of the nearest particle which is already 
on the surface.
This probability is given by
\begin{equation}
p(r) = 
\left\{
\begin{array}{ll}
1              &  r \le d          \\
(d/r)^{\alpha} & r>d,
\end{array}
\right.
\label{stick.prob}
\end{equation}
where the exponent $\alpha \ge 0$ 
is a parameter of the model.
The model thus exhibits a positive feedback clustering, like
many of the clustering phenomena listed above.
The random deposition process is repeated until the desired
number of particles, $M$, stick to the surface.
Since the sticking probability is given by a power law function,
which is of a long range nature, this process should have been 
studied, 
in principle,
in the infinite system limit.
Numerical simulations, however, are done on a finite system
of area $L \times L$, where $L/d \gg 1$. 
The coverage is given by 
$\eta = A/L^2$
where $A$ is the total area covered by the particles. Also, let
$\eta_0 =  \pi (d/2)^2 \cdot M/L^2$.

The limit of {\it uncorrelated adsorption}, 
in which
the sticking probability is
$p=1$ uniformly,
is obtained
for $\alpha=0$. 
This limit was studied recently using fractal
analysis, and the box-counting and Minkowski functions were calculated 
analytically 
\cite{Hamburger96a}.
It was found that for a range of low coverages, apparent fractal
behavior is observed between physically
relevant cutoffs.
The lower cutoff $r_0$ is given by the particle diameter $r_0=d$ while the
upper cutoff $r_1$ is given by the average gap between adjacent 
particles, namely
$r_1 = \rho^{-1/2} - d$,
where $\rho = M/L^2$ is the particle density.

The limit of {\it strongly correlated adsorption} is obtained 
for $\alpha \rightarrow \infty$.
In this case only a single, connected cluster is generated on 
the surface. 
The perimeter of this cluster grows slowly when
new particles are deposited on its edge, while it becomes more
dense inside
\cite{Provatas95,Provatas97}.

We will now examine
the morphological properties of the 
configurations of adsorbed particles 
for the full range
of $0<\alpha<\infty$
using fractal analysis.
For this analysis we use the box-counting (BC) procedure 
in which one covers the plane by
non-overlapping boxes of linear size $r$.
The box-counting function
$N(r)$
is obtained by counting the number of boxes
which have a non-empty intersection with the (fractal) set.  
A fractal dimension $D_{BC}$, is declared to prevail 
at a certain range of length-scales, if a relation
of the type
$N_{BC}(r) \sim r^{-D_{BC}}$
holds, or equivalently, 
if the slope of the log-log plot

\begin{equation}
D_{BC} = - {\rm slope}\:\{\log r, \log[N_{BC}(r)] \}
\label{eq:D_BC}
\end{equation}

\noindent 
is found to be constant over that range.

Two configurations of particles,
randomly deposited and adsorbed according to
Eq.
(\ref{stick.prob})
onto the unit square ($L=1$),
are shown in Fig. 1 for
$\eta_0=0.01$.
For 
$\alpha=1.5$  
the particle distribution 
exhibits local density fluctuations
but on larger scales it is rather homogeneous
and extends over
the entire system
[Fig. 1(a)].
For
$\alpha=2.5$  
we observe a strongly clustered structure 
[Fig. 1(b)].
This structure resembles the set of turning points of
a L\'evy flight random walker 
\cite{Shlesinger94}.
In fact, a L\'evy flight corresponds to the special case 
in which the sticking probability of the next 
deposited particle depends only on the position 
of the latest particle adsorbed
on the surface. 
Unlike  L\'evy flights which typically describe dynamic behavior 
our model describes clustering in spatial structures.
It is also related to other models of spatial structures
such as the continuous percolation model which is approached 
when the interaction is suppressed, at $\alpha \rightarrow 0$.
Another related model, which describes the growth of a percolation
cluster and exhibits power-law correlations between growth sites
is presented in Ref.
\cite{Bunde85}.

The box-counting functions for the configurations 
generated by the PLSA model 
are shown in Fig. 2. 
It is observed that for $\alpha < 2$ the box-counting
function resembles the shape obtained for the uncorrelated
case
\cite{Hamburger96a}.
This indicates that the basic features of the model studied
in Ref. 
\cite{Hamburger96a}
are maintained not only for short range correlations
but for an entire class of long range correlations.
The box-counting function for 
$\alpha>2$ exhibits a nearly linear
behavior for the entire range from 
the particle size to the cluster size.

To obtain the fractal dimensions of the sets 
from the box-counting functions
one 
should identify the relevant range of length-scales over
which the linear fit should be done.
For the weakly correlated distributions 
the relevant range of length scales spans between
$r_0=d$ and 
$r_1= \rho^{-1/2} - d$
\cite{Hamburger96a}.
For the strongly correlated distributions 
where clusters are formed,
the relevant range is limited from above 
by the linear size of the entire cluster. 
The quality of the linear fit is measured by
the coefficient of determination ${\cal R}^2$
\cite{Hamburger96a}.
In both cases, given a desired value of
${\cal R}^2$
one can further narrow the range within the cutoffs described 
above to find the broadest range $(r_0,r_1)$
within which the linear regression maintains the 
given value of ${\cal R}^2$
\cite{ranges}.

The fractal dimension $D$ as a function of 
$\alpha$ is shown in Fig. 3
for $\eta_0=0.1$, $0.01$ and $0.001$.
Two domains are identified: 
a plateau of low dimension for the
weakly correlated case
and a plateau of high dimension for the strongly
correlated case. 

Consider a seed particle located at the origin 
of an infinite plane.
Particles are randomly deposited one at a time 
according to the PLSA model
until one particle sticks to the surface. 
Consider the probability that the distance
$r$ between the first particle which sticks and the
seed particle at the origin will be larger than some 
value $r_f$ where 
$r_f>d$. 
This probability is given by:

\begin{equation}
P(r>r_f) = { {\int_{r_f}^{\infty} 2 \pi r ({d \over r})^{\alpha} dr}
\over
{\int_{0}^{d} 2 \pi r dr +
\int_{d}^{\infty} 2 \pi r ({d \over r})^{\alpha} dr} }.
\end{equation}
One readily verifies that for
$\alpha<2$ 
the probability
$P(r>r_f)=1$ for any finite $r_f$.
For 
$\alpha>2$, on the other hand,
this probability is given by
\begin{equation}
P(r>r_f)= {2 \over \alpha} \left({d \over r_f}\right)^{\alpha-2}. 
\label{prob.decay}
\end{equation}

\noindent
Therefore, in the infinite system limit of the weakly correlated phase 
($\alpha<\alpha_c$)
the probability that the next particle will stick within any finite distance
from an existing cluster is zero
\cite{comment.fibers}.
In the strongly correlated phase
($\alpha>\alpha_c$),
the probability that
the next particle will stick within a finite distance 
$r_f$ from the cluster can be made arbitrarily close to one,
by an adjustment of $r_f$ according to Eq. 
(\ref{prob.decay})
\cite{comment.exp}.
In general, the value of
$\alpha_c$ 
for which the clustering transition takes place
is equal to the space dimension.

The order parameter of the clustering transition is 
$V = (\eta_0 - \eta)/\eta_0$,
namely the fraction of the total area of the adsorbed particles
which is lost due to overlap.
Consider a finite number $M$ of particles of diameter $d=1$
adsorbed on the surface
in the infinite system limit 
$(L \rightarrow \infty)$.
For $\alpha< 2$, in this low coverage limit overlaps are
negligible and $V=0$.
For $\alpha>2$ clusters become more dense and overlaps more dominant
as $\alpha$ increases.
Our numerical studies are done on a finite system of size 
$L=1$ for a range of coverages.
The order parameter $V$ as a function of $\alpha$,
for $\eta_0=0.01$,
is shown in Fig. 4.

To examine the critical behavior in the infinite system limit we
performed analytical calculations in one dimension (1D).
In 1D the configuration
is fully specified by the ordered list of $M-1$ gaps between
the $M$ particles. 
For the 1D case we have obtained the critical exponent 
$\beta$ for the order parameter
$V \sim (\alpha - \alpha_c)^{\beta}$
in the $L \rightarrow \infty$ limit by constructing the probability distribution
$P_i$, $i=0,\dots,M$ that the next particle that sticks will stick within the
gap $g_i$ (where $g_0$ and $g_M$ are the two semi-infinite gaps on both sides).
The overlap was then calculated as a weighted average over all gaps.
The result we obtain is that the exponent $\beta = 1$. 
We also found 
that the fractal
dimension exhibits critical behavior of the form 
$D \sim (\alpha - \alpha_c)^{\gamma}$, where $\alpha_c = 1$
with the exponent $\gamma=1$.
For the weakly correlated phase at $\alpha < 1$ the fractal dimension in
the $L \rightarrow \infty$ is $D=0$ while for $\alpha>2$ the dimension is
$D=2$. 

In summary, we introduced a new model for random sequential 
adsorption, characterized by a power law distribution of sticking
probabilities. 
This model exhibits a continuous phase transition between weakly
correlated adsorption, 
in which
the particle distribution 
is homogeneous
on large scales
and extends over
the entire system,
and strongly correlated  
adsorption in which 
a highly clustered structure is generated. 
We thus identified a
broad class of distributions which maintain the basic properties of the
weakly correlated random structures studied in Ref. 
\cite{Hamburger96a}
and found the borderline between this class and the class of strongly
correlated structures which exhibit clustering phenomena.
The model should be useful in the study of a great variety of
clustering problems.

\onecolumn

\begin{figure}
\caption{Particles adsorbed on the surface of a unit square $(L=1)$ 
according to the PLSA model for 
$\eta_0=0.01$: 
(a) for $\alpha=1.5$ we observe density fluctuations at small scales, however
at larger scales the distribution is rather homogeneous and extends over the
entire system; 
(b) for $\alpha=2.5$ we observe a strongly clustered structure and vacant area
elsewhere. 
The number of particles in both (a) and (b) is 3184
and their diameter is 
$d = 0.002$.
}
\label{fig1}
\end{figure}

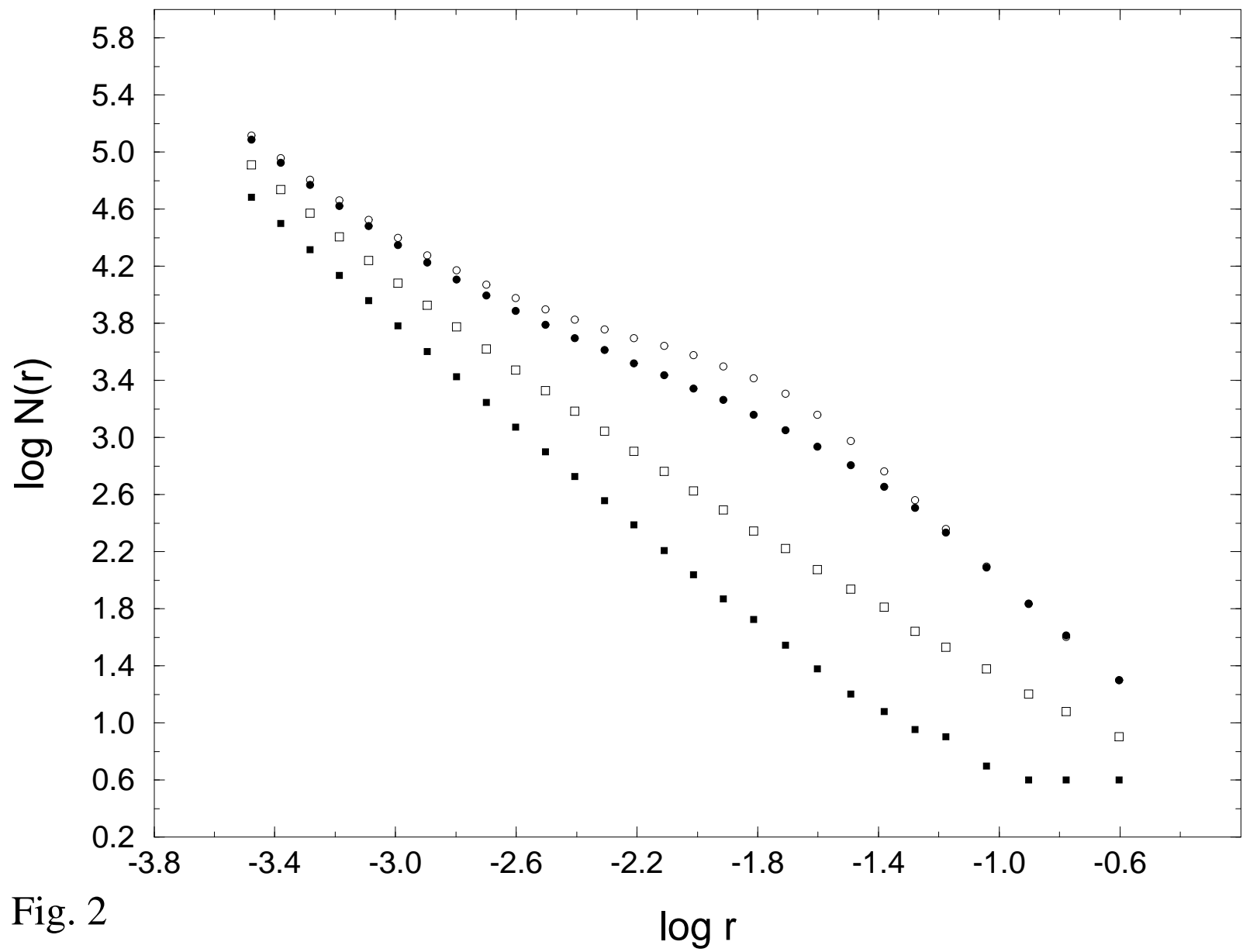
\begin{figure}
\caption{The box-counting function for four configurations with
$\eta_0=0.01$, and $\alpha=0$ (empty circles), $1.5$ (full circles),
$2.5$ (empty squares) and $3.5$ (full squares).
It is observed that for $\alpha<2$ the shape of this function 
resembles that of the uncorrelated case.
For $\alpha>2$, where strongly clustered distributions arise,
there is a broad scaling range. The units are dimensionless and
the logarithms are in base 10.
}
\label{fig2}
\end{figure}

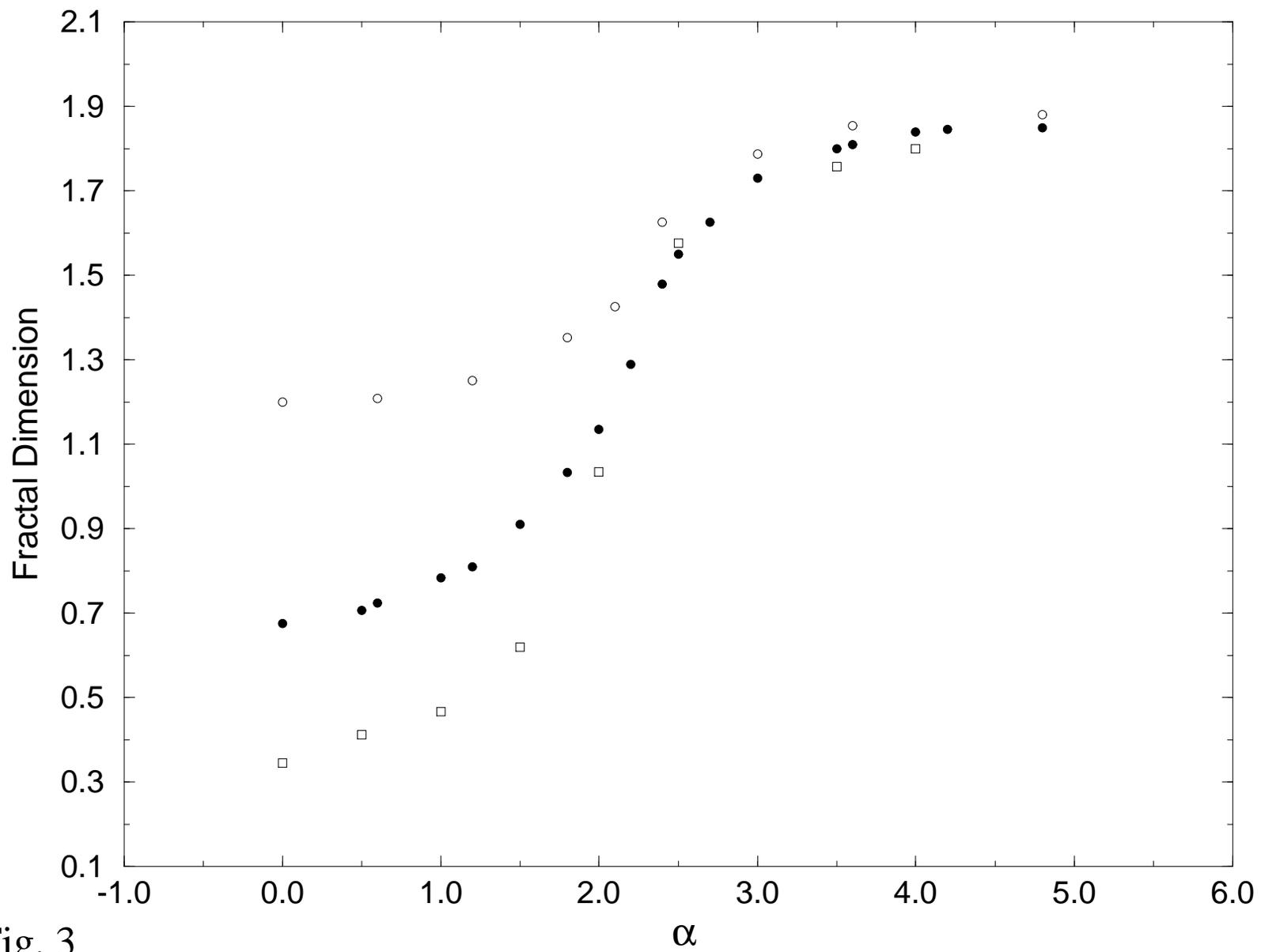
\begin{figure}
\caption{The fractal dimension of the configurations produced by the
PLSA model
as a function of $\alpha$ (in dimensionless units)
for $\eta_0=0.1$ (empty circles), 
$0.01$ (full circles)
and $0.001$ (empty squares).
A plateau of low dimension is found for the weakly correlated
limit and a plateau of high dimension for the
strongly correlated limit. 
}
\label{fig3}
\end{figure}

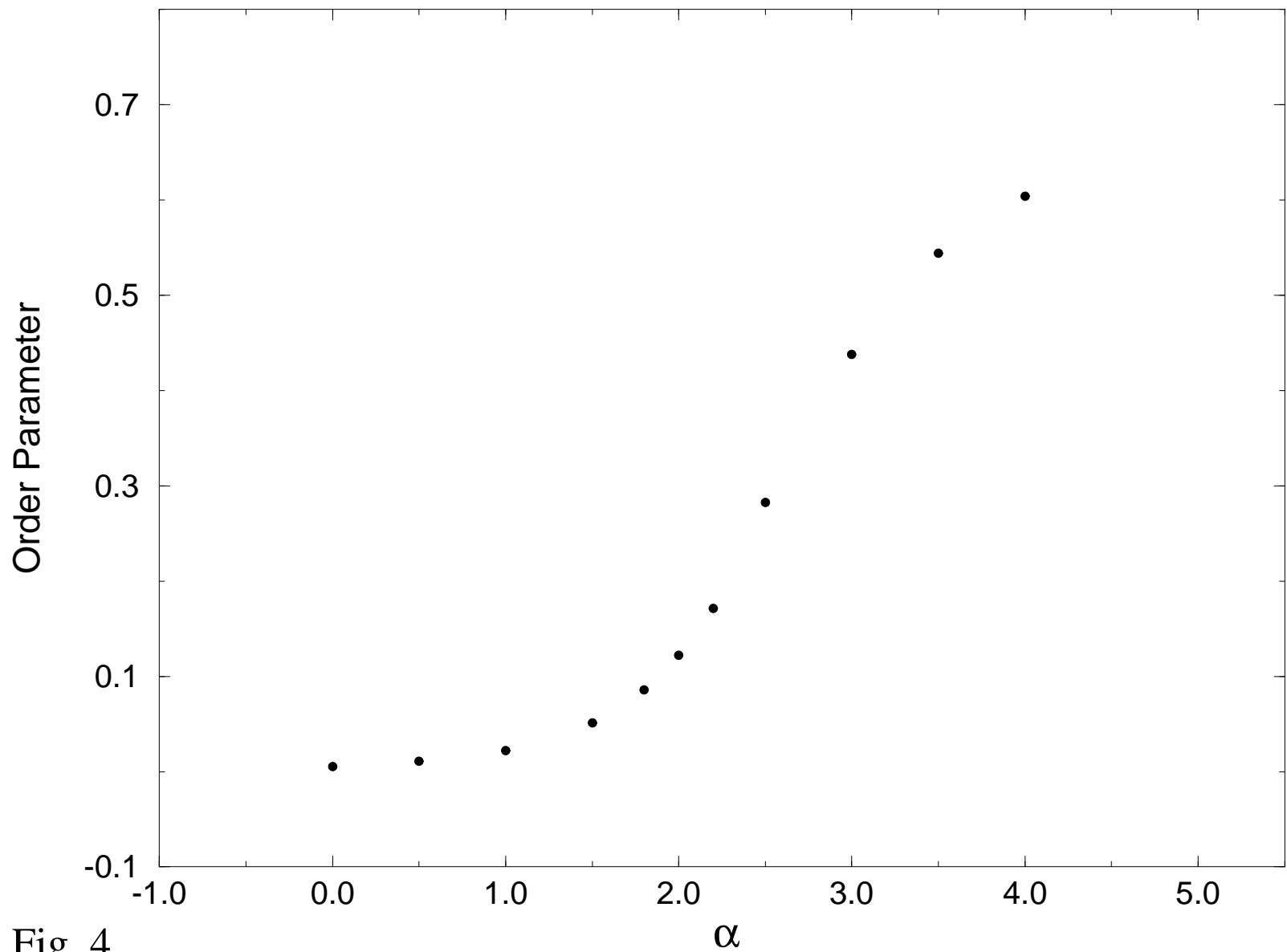
\begin{figure}
\caption{
The order parameter of the clustering transition,  
$V = (\eta_0 - \eta)/\eta_0$,
is shown as a function of $\alpha$
for $\eta_0=0.01$.
It represents the
fraction of the total area of the adsorbed particles
which is lost due to overlap.
For $\eta_0 \ll 1$, this order parameter vanishes for $\alpha<2$
and increases above $\alpha=2$. 
}
\label{fig4}
\end{figure}


\begin{thebibliography}{10}
\bibitem{ranges}
For the linear regression of the box-counting function in Fig. 2 
we chose ${\cal R}^2=0.996$ 
and obtained that for
$\alpha=0$, $\log r_0=-2.7$ and $\log r_1=-1.8$.
For
$\alpha=2.5$, the same value of 
${\cal R}^2$ is obtained for a broader range
between $\log r_0=-3.5$ and $\log r_1=-0.8$.
Note that in the latter case $\log r_0$ is somewhat
smaller than $\log d$ since the gaps between
particles in the cluster can be smaller than $d$.
\bibitem{comment.fibers}
{Note that the fiber deposition model of Ref. \cite{Provatas97} with sticking
  coefficient $P(r)=1$ for $r \le d$ and $P(r)=p$ for $r>d$ shares this
  property for any $0<p \le 1$.}
\bibitem{comment.exp}
{Note that the class of models which give rise to strong clustering is broad
  and includes all models in which $p(r)$ decays faster than a power-law. For
  example, a model with $P(r)=1$ for $r \le d$ and $P(r) = e^{-k(r-d)}$ for
  $r>d$ also gives rise to strong clustering for any $k>0$.}
\end{thebibliography}
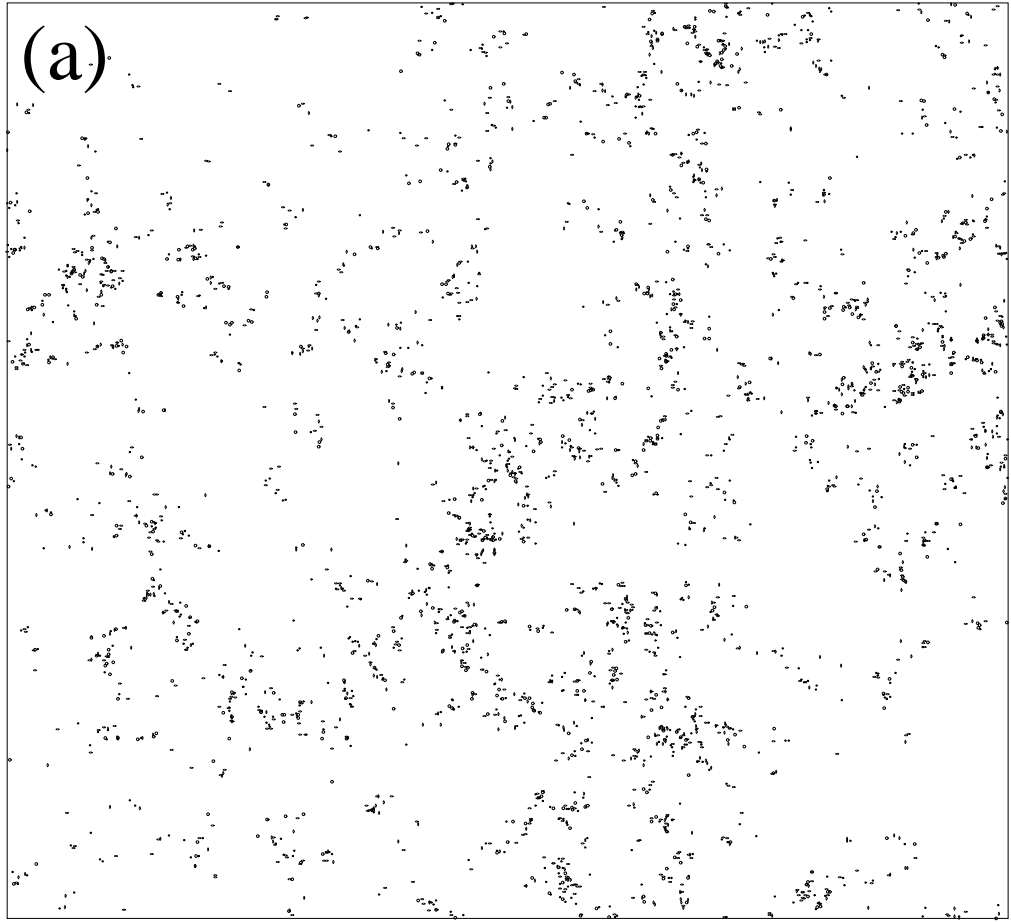
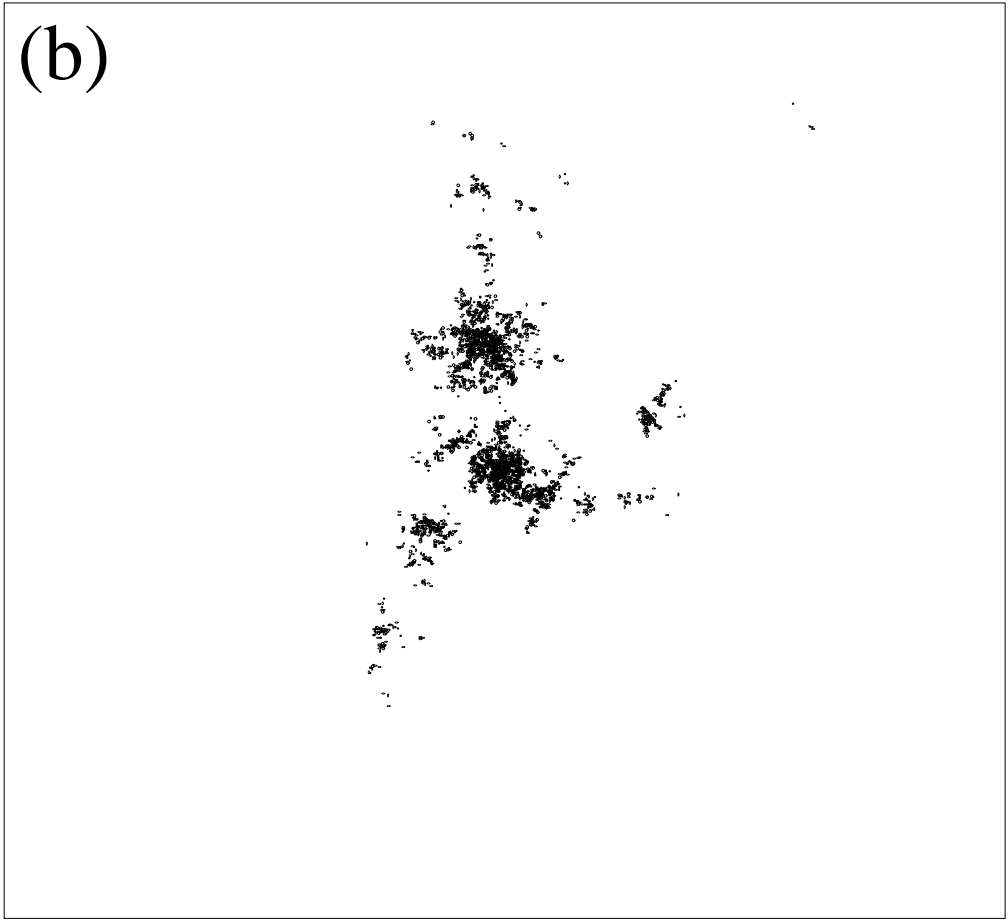
\end{document}